# Clustering Fetal Heart Rate Tracings by Compression

Costa Santos, C,
*Biostatistics and Medical Informatics Department and Centre for Research in Health Information Systems and Technologies, Faculty of Medicine, University of Porto, Portugal.*
csantos@med.up.pt

Bernardes, J.
*Obstetrics and Gynaecology Department. Faculty of Medicine, University of Porto, Portugal.*
jbernardes@mail.telepac.pt.

Vitányi, P.M.B.
*Centre for Mathematics and Computer Science (CWI). Computer Science Department, Faculty of Science, University of Amsterdam, The Netherlands.*
Paul.Vitanyi@cwi.nl

Antunes, L.
*Computer Science Department. Artificial Intelligence and Computer Science Lab. Faculty of Science, University of Porto, Portugal.*
lfa@ncc.up.pt

## Abstract

*Fetal heart rate (FHR) monitoring is widely used regarding the detection of fetuses in danger of death or damage. Thirty one FHR tracings acquired in the antepartum period were clustered by compression in order to identify abnormal ones. A recently introduced approach based on algorithmic information theory was used. The new method can mine patterns in completely different areas, without domain-specific parameters to set, and does not require specific background knowledge. At the highest level the FHR tracings were clustered according to an unanticipated feature, namely the technology used in signal acquisition. At the lower levels all tracings with abnormal or suspicious patterns were clustered together, independently of the technology used.*

## 1. Introduction

Cardiotocography is widely used, all over the world, for fetal heart rate (FHR) and uterine contractions monitoring before (antepartum) and during labor (intrapartum), regarding the detection of fetuses in danger of death or permanent damage [1]. However, more than 30 years after its introduction into clinical practice, the predictive capability of the method remains controversial [2]. Indeed, although several studies addressed this issue, they are incomparable because they were performed in different conditions, using different equipment, different interpretation criteria, different definitions of poor neonatal outcome and different time intervals that elapse between the end of obtaining the tracings and the evaluation of neonatal state at delivery [3]. Moreover, intra and inter-observer variation in visual analysis FHR tracings remains a large and unsolved issue [4,5].

Computer analysis of cardiotograms provides quantitative parameters that are difficult to assess by the human eye, and overcomes observer variation in FHR analysis. SisPorto 2.0 is a program for automated analysis of tracings, developed over the last 15 years at the University of Porto [6,7]. SisPorto defines FHR baseline using a complex algorithm developed to identify the mean FHR during stable segments, in the absence of fetal movements and uterine contractions. Accelerations are defined as the increases in the FHR above baseline, lasting 15–120 seconds and reaching a peak of at least 15 beats per minute (bpm). Accelerations are defined as decreases in the FHR under the baseline, lasting at least 15 seconds and with amplitude exceeding 15 bpm. They are classified as mild if shorter than 120 seconds, prolonged if they last 120–300 seconds, and severe if they exceed 300 s. Points with abnormal short-term-variability (STV) are recognized when the difference to adjacent FHR signals is less than 1 bpm. Points with abnormal long-term variability (LTV) are identified whenever the difference between maximum and minimum FHR values of a sliding 60 seconds window centered on them, does not exceed 5 bpm [7]. Computerized quantification of accelerations and abnormal STV and LTV with SisPorto allows good prediction of 1 minute and 5 minutes Apgar scores [8], a method, to evaluate a newborn's adjustment to extra uterine life. Apgar score ranges from 0 to 10. A low score represents neonatal distress whereas a high score indicates absence of difficulty in adjusting to extra uterine life.

In 2004/2005, a new method for data clustering using compression was introduced by Cilibrasi & Vitányi [9], based on a similar method first used in mitochondrial genome phylogeny [10]. It is feature-free, there are no parameters to tune, and no domain-specific knowledge is used. Good results where obtained applying it on different areas: languages tree, genomics, optical character recognition, literature [9], music [11], and computer virus and internet traffic analysis [12]. The algorithm goes in two steps: first it determines a universal similarity distance between each pair of objects, computed from the length of compressed data files. Secondly, a hierarchical clustering method is applied.

The aim of this study was to cluster FHR tracings by this compression method in order to identify and classify abnormal FHR tracings with different, distinguishing, patterns.

## 2. Clustering Based on Compression

The new approach is based on so-called algorithmic information theory, a theoretical, rigorous and well studied notion of information content in individual objects [9]. The method is extremely powerful and general. It can mine patterns in completely different areas, there are no domain-specific parameters to set, and it does not require any background knowledge. Its performance in clustering heterogeneous data and anomaly detection in time sequences has been shown to outperform that of every known data-mining method in a massive recent study [13]. In this study, we applied the method directly to current clinical data rather than on benchmark data sets. This may be the first application of this type for the compression based method. We used the implementation in the CompLearn package [14], which determines the distances and computes the tree from the digitized tracings. For a fixed given compressor, like gzip, bzip2, or PPMZ, we denote the size in bits of the compressed version of a file x by C(x). Given a set of FHR tracings, we compresse them singly, noting the file size of each compressed version. Then, for every pair of FHR tracings, say x and y, we append the two of them to a single file xy, and compress it, yielding a C(xy) bit file. Subsequently, we compute the difference between the length of the compressed file of the pair of FHR traces and the minimum of the lengths C{x}, C{y}, of the compressed versions of the two constituent FHR tracings. Finally, we divide this difference by the maximum of the lengths C(x), C(y) of the compressed versions of the two constituent FHR tracings, in order to normalize the values to be between 0 and 1, so that relative comparison between instances is possible. This yield the `normalized compression distance' or `NCD' between x and y:

$$NCD(x,y) = \frac{C(xy) - \min\{C(x), C(y)\}}{\max\{C(x), C(y)\}}.$$

Doing this for all pairs of tracings we yield a 2-dimensional distance matrix (the entries are the pair wise NCD distances between the FHR tracings). The smaller the above difference for a pair of tracings, the more likely they are to be the same, i.e., if two FHR are similar, then we can succinctly describe one given the other. This happens because the compressor looks for patterns in the FHR trace to shrink it without losing information. This approach is universal in the sense that it aims at discovering all effective similarities, [9], instead of using some particular features to cluster. This is the theory in practice 'all' means 'a large family related to the compressor involved'. Given a set of objects the pairwise NCD's form the entries of a distance matrix. Just as the distance matrix is a reduced form of information representing the original data set, we need to reduce the information even further in order to achieve a cognitively acceptable format like data clusters. The distance matrix contains all the information in a raw form that is not easily usable, since for n>3 our cognitive capabilities rapidly fail. In our situation we do not know the number of clusters a-priori, and we let the data decide the clusters. The most natural way to do so is hierarchical clustering [15]. Such methods have been extensively investigated in Computational Biology in the context of producing phylogenies of species.

One the most the sensitive ways is to exactly resolve a bifurcating tree given weighted so-called 'quartet topologies'. This method is sensitive, but NP-hard and hence infeasible in practice. Therefore, we use a new heuristic that monotonically approximates the global optimum. It is time consuming, running in quartic time. Other inexact hierarchical clustering heuristics, like parsimony, may be much faster, quadratic time, but they are less sensitive. Non exact incremental quartet methods like the popular quartet-puzzling [16] run faster than our method but yield non optimal trees with agreement values on the branches. As current compressors are good but limited, we want to exploit the smallest differences in distances, and therefore use the most sensitive method to get the greatest accuracy.

In this study, we used a new exact quartet-method, which is a heuristic based on randomized parallel hill-climbing genetic programming. It monotonically approximates the global optimum and delivers the single tree that approximates the global optimum best at termination. A detailed description of this method is presented elsewhere [9,17]. To cluster n data items, the algorithm generates a random ternary tree with n-2 internal nodes and n leaves. The algorithm tries to improve the solution at each step by interchanging sub-trees rooted at internal nodes (possibly leaves). It switches if the total tree cost is improved. To find the optimal tree is NP-hard, that is, it is infeasible in general. To avoid getting stuck in a local optimum, the method executes sequences of elementary mutations in a single step. The length of the sequence is drawn from a fat tail distribution, to ensure that the probability of drawing a longer sequence is still significant. In contrast to other methods, this guarantees that, at least theoretically, in the long run a global optimum is achieved. Because the problem is NP-hard, we can not expect the global optimum to be reached in a feasible time in general. Yet for natural data, as in this study, experience shows that the method usually reaches an apparently global optimum. One way to make this more likely is to run several optimization computations in parallel, and terminate only when they all agree on the solutions (the probability that this would arise by chance is very low as for a similar technique in Markov chains). The method is so much improved against previous exact quartet-tree methods, that it can cluster larger groups of objects (over 100) than was previously possible (around 15-30). Other, inexact, methods like quartet-puzzling construct incremental solutions. It is almost guaranteed that such methods cannot reach a global optimum.

Our clustering heuristic generates a tree with a certain fidelity with respect to the underlying distance matrix (or alternative data from which the quartet tree is constructed) called standardized benefit score or S(T) value in the sequel. This value measures the quality of the tree representation of the overall order relations between the distances in the matrix. It measures in how far the tree can represent the quantitative distance relations in a topological qualitative manner without violating relative order. The way it works is briefly as follows. Every quartet, consisting of 4 elements x,y,z,w, out of the n elements, can be arranged in 3 quartet topologies denoted as xy|zw, xz|yw, xw|yz. Every such topology has a weight: the weight of xy|zw equals NCD(x,y)+NCD(z,w). A bifurcating tree embeds each quartet in the form of one of the three topologies. To optimize the tree we want to find the tree such that the sum of the weights of all embedded quartet topologies is minimal. The best we can do is to embed every quartet at minimal weight, call the total sum of weights m. The worst we can do is to embed every quartet at maximal weight, call the total sum of weights M. Every bifurcating tree T with the n elements as leaves will have a sum C(T) of the weights of the embedded quartet topologies somewhere in between.

S(T)=(M-C(T))/(M-m), where the S(T) value ranges from 0 (worst) to 1 (best). A random tree is likely to have S(T) about 1/3, while S(T)=1 means that the relations in the distance matrix are perfectly represented by the tree. Since we deal with n natural data objects, living in a space of unknown metric, we know a priori only that the pairwise distances between them can be truthfully represented in (n-1)-dimensional Euclidean space. Multidimensional scaling, representing the data by points in 2-dimensional space, most likely necessarily distorts the pairwise distances. This is akin to the distortion arising when we map spherical earth geography on a flat map. A similar thing happens if we represent the n-dimensional distance matrix by a ternary tree. It can be shown that some 5-dimensional distance matrices can only be mapped in a ternary tree with S(T)<0.8. Practice shows, however, that up to 12-dimensional distance matrices, arising from natural data, can be mapped into a such tree with very little distortion S(T)>0.95. In general the S(T) value deteriorates for large sets. The reason is that, with increasing size of natural data set, the projection of the information in the distance matrix into a ternary tree gets necessarily increasingly distorted. If for a large data set like in this paper, 31 objects, the S(T) value is as large as here, S(T)=0.944, then this gives evidence that the tree faithfully represents the distance matrix, but also that the natural relations between this large set of data were such that they could be represented by such a tree.

## 3. Fetal Heart Rate Tracings

Thirty one FHR tracings, acquired in the antepartum period, in 4 hospitals, renamed here as H1, H2, H3 and H4, within a multicentre observational study for SisPorto validation, were used for clustering [8].

Tracings acquisition with SisPorto were performed from Hewlett-Packard M1350 fetal monitors at a 4 Hz sampling rate (4 samples per second), in 3 centers, and with a Sonicaid 8000 fetal monitor, in the other centre, at a 3 Hz (3 samples per second). All tracings had at least 30 minutes duration and less than 15% signal loss. Multiple gestations and fetal malformations were not included, and only FHR tracings acquired within 4 hours preceding delivery, were included. In each FHR tracing the following parameters were extracted from the SisPorto analysis, as described before: FHR baseline, number of accelerations, percentage of tracing with abnormal STV and LTV and average STV and LTV. Apgar scores at minutes 1 and 5 were evaluated by the health caregivers responsible for immediate neonatal support

## 4. Results

In Figure 1, we clustered the data from Table 1, obtaining a representation with a high S(T) value, 0.944.

**Table 1.** Tracings description: tracing code (T), hospital code (H), neonatal outcome (Apgar at minutes 1 and 5) and parameters extracted by SisPorto: tracing duration in minutes (D); baseline (BL); number of accelerations (Acc); % of abnormal (% abn) and average (av) of STV and LTV.

| T-H | Apgar 1 | Apgar 5 | D | BL | Acc | STV % abn | STV av | LTV % abn | LTV av |
|---|---|---|---|---|---|---|---|---|---|
| A | 1 | 4 | 6 | 43 | 131 | 0 | 76 | 0,4 | 33 | 7,7 |
| B | 1 | 4 | 7 | 52 | 125 | 0 | 86 | 0,4 | 38 | 8 |
| C | 1 | 5 | 8 | 45 | 151 | 0 | 78 | 0,6 | 46 | 7,4 |
| D | 1 | 6 | 9 | 60 | 156 | 0 | 75 | 0,7 | 62 | 5,5 |
| E | 1 | 1 | 5 | 40 | 123 | 0 | 88 | 0,3 | 71 | 5,7 |
| F | 1 | 2 | 7 | 43 | 135 | 0 | 85 | 0,6 | 75 | 5 |
| Y | 2 | 9 | 10 | 40 | 135 | 0 | 72 | 0,8 | 23 | 7,2 |
| Zc | 1 | 5 | 8 | 60 | 141 | 0 | 75 | 1,1 | 12 | 14,3 |
| T | 1 | 9 | 10 | 43 | 122 | 26 | 54 | 2,1 | 0 | 11,3 |
| U | 1 | 9 | 10 | 30 | 121 | 9 | 61 | 1,6 | 0 | 7 |
| H | 1 | 9 | 10 | 32 | 134 | 22 | 59 | 0,7 | 0 | 5,1 |
| I | 1 | 7 | 8 | 54 | 124 | 17 | 65 | 0,9 | 0 | 9,3 |
| J | 1 | 8 | 9 | 57 | 131 | 10 | 64 | 1,5 | 0 | 10,6 |
| K | 1 | 9 | 10 | 50 | 125 | 22 | 62 | 1,4 | 0 | 9,2 |
| L | 1 | 7 | 9 | 54 | 126 | 17 | 61 | 1,5 | 0 | 8,7 |
| Za | 1 | 6 | 9 | 40 | 134 | 8 | 63 | 0,7 | 18 | 6,8 |
| M | 2 | 9 | 10 | 41 | 110 | 11 | 47 | 4 | 0 | 3,1 |
| N | 3 | 10 | 10 | 46 | 113 | 30 | 48 | 2,4 | 0 | 5,4 |
| P | 2 | 9 | 10 | 40 | 120 | 31 | 49 | 2,4 | 0 | 3,9 |
| Q | 3 | 10 | 10 | 50 | 110 | 42 | 49 | 3,2 | 0 | 1,2 |
| Zh | 4 | 6 | 10 | 60 | 110 | 28 | 54 | 0,9 | 0 | 6,7 |
| O | 2 | 9 | 10 | 40 | 134 | 25 | 49 | 3,2 | 0 | 4,6 |
| R | 4 | 5 | 8 | 44 | 128 | 15 | 50 | 2,2 | 0 | 7,9 |
| S | 4 | 9 | 9 | 45 | 129 | 24 | 53 | 2,2 | 0 | 8,2 |
| V | 3 | 9 | 10 | 58 | 138 | 15 | 61 | 1,3 | 0 | 10 |
| Ze | 3 | 9 | 10 | 60 | 135 | 47 | 55 | 2 | 0 | 2,9 |
| Zf | 3 | 10 | 10 | 46 | 125 | 39 | 55 | 4,1 | 0 | 0,9 |
| Zg | 4 | 9 | 9 | 50 | 125 | 31 | 54 | 2,5 | 0 | 5 |
| X | 4 | 10 | 10 | 47 | 129 | 14 | 64 | 1 | 19 | 7,9 |
| Z | 2 | 9 | 10 | 44 | 126 | 21 | 59 | 1 | 22 | 1 |
| Zb | 3 | 9 | 9 | 37 | 142 | 0 | 63 | 1,1 | 17 | 8,4 |

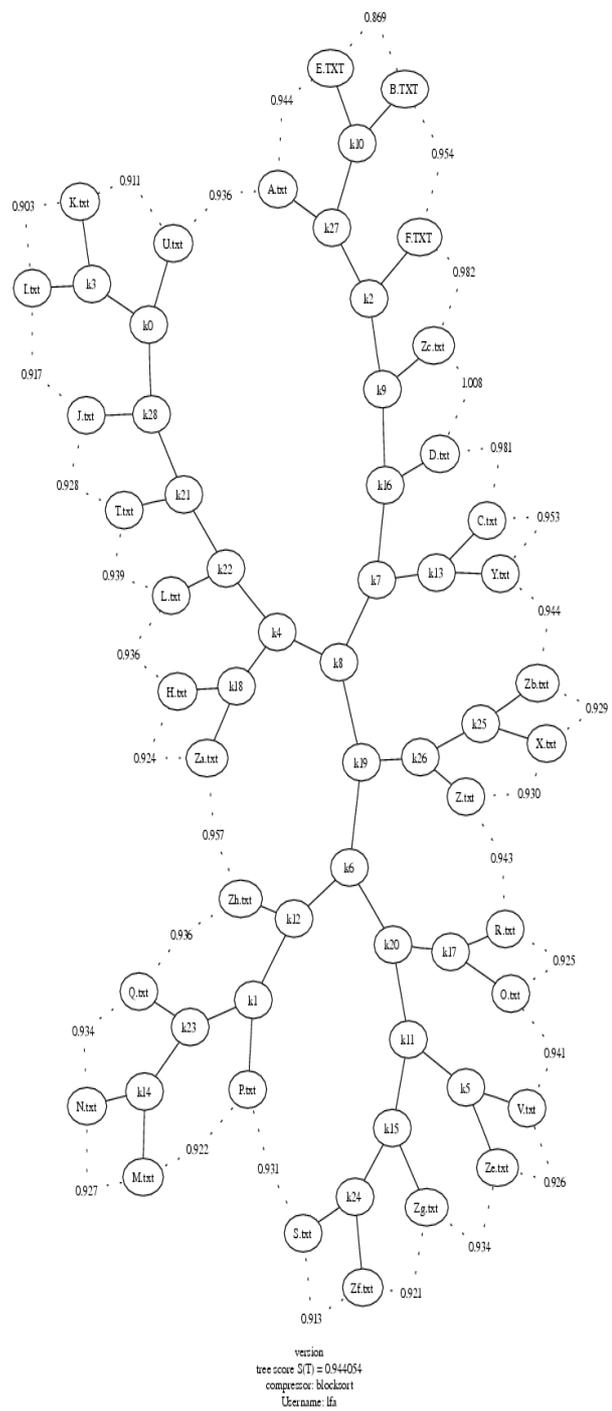

**Figure 1.** Hierarchical clustering of FHR data according to NCD distances.

This means that the tree represents the relations between the data in the distance matrix faithfully. Because it is this high for such a large group of data this gives also evidence that the NCD distances involved captured properties of the data which naturally induced such a hierarchical tree structure.

That is, we can have some confidence that the tree not only depicts the NCD distance matrix faithfully, but also that the similarity relations between the FHR tracings were naturally treelike with this hierarchical clustering. Since there are different ways to lay out the same tree on a 2-dimensional plane, the CompLearn package looks for the most natural way. This is interpreted as the way that minimizes the sum of the NCD's between the ordered sequences of leaves on the convex hull of the tree layout (the dotted line in Figure 1). This is the meaning of the numbers on the dotted line around the tree: the NCD's between the adjacent leaves in the optimal convex hull label each segment.

Interpreting Figure 1 and Table 1 we found the following clustering: as a first grouping, tracings where split into tracings obtained at hospital H1 and tracings obtained at hospitals H2, H3 and H4. Going back to the original data, and conditions under which they were acquired, we found that signal acquisition was performed by different technologies in H1 and other hospitals. At H1 signals were acquired on average 3 times/second whereas in other hospitals on average 4 times/second. This division in itself is of course not very interesting but shows that the method detects this situation. Looking closer in the clusters we found four clear clusters: suspicious/pathologic, normal from H1 hospital, normal from hospitals H2, H3 and H4 and other tracings.

**4.1. Suspicious/pathologic cluster**

This cluster with both hospital H1 and hospital H2 tracings contains all cases (A, B, C, D, E, F, Y and ZC) with high percentage of abnormal STV (>70%), all other tracings have abnormal STV≤65% and we know that high percentage is a good predictor for poor neonatal outcome [8]. In fact all tracings with poor neonatal outcome, Apgar score at first minute lower than 5, were included in this cluster.

**4.2. Normal H1 cluster**

This cluster with only tracings from hospital H1 contains all cases (T, U, H, I, J, K, L and ZA) with percentage of abnormal STV<65%. These tracings also have lower percentage of abnormal LTV (median=0, range from 0 to 18) than tracings from suspicious/pathologic cluster (median=42, range from 12 to 75), and more accelerations (median=17, range from 8 to 26) than tracings from suspicious/pathologic cluster (median=0, range from 0 to 0). Both low percentage of abnormal LTV and the existence of acceleration are associated with a good neonatal outcome [8].

**4.3. Normal H2, H3 and H4 cluster**

This cluster with only tracings from hospitals H2, H3 and H4 contains all cases (M, N, P, Q, ZH, O, R, S, V, ZE, ZF and ZG) with percentage of abnormal STV<65%. These tracings also have lower percentage of abnormal LTV (median=0, range from 0 to 0) than tracings from suspicious/pathologic cluster (median=42, range from 12 to 75), and more accelerations (median=29, range from 11 to 47) than tracings from suspicious/pathologic cluster (median=0, range from 0 to 0). Both low percentage of abnormal LTV and the existence of acceleration are associated with a good neonatal outcome [8]. This cluster was split into two sub-clusters: one with lower baseline, range from 110 to 120, and other with baseline higher than 125.

**4.4. Other tracings cluster**

Finally, there is a cluster with 3 tracings (X, Z and ZB). These tracings do not belong to suspicious/pathologic cluster because the percentage of abnormal STV ranges from 59 to 64, but they have higher abnormal LTV percentage than tracings from other centers cluster and they belong to hospitals H1, H2 and H3.

## 5. Discussion and Future Work

In our analysis, FHR tracings were clustered in the highest level by an unexpected feature: the technology used in signal acquisition. However, all tracings with high percentages of abnormal STV and LTV and no acceleration are clustered together in the lower-level clusters, independently from the technology used in signal acquisition, and these parameters are good predictors of a poor neonatal outcome [8]. Although the value of the Apgar score as an indicator of neonatal outcome has been questioned, because the majority of newborns with low estimates are known to have a normal post-natal course it is almost certain that in the group of newborns with low Apgar scores are included all of those who will die or develop severe handicaps as a consequence of peri-partum fetal hypoxia.

Our results, that need confirmation in larger studies, compare favorably with the results from the conventional methods of FHR monitoring reported in international guidelines, considering the analysis of FHR accelerations and variability [2,18], as can be inferred from Table 1.

Our analysis may be improved with more tracings, with good and poor outcome, but this is not easy because low incidence of newborns with poor neonatal outcome is a current reality in industrialized countries. Moreover, as the ultimate goal of clustering FHR tracings is their classification as suspicious/ pathological or normal, this can be done using the SVM-NCD approach published in 2004 [19]. In an initial phase hierarchical cluster should be used to determine the set of k classes we want to distinguish in the sequel. Subsequently, a k-ary SVM-NCD classifier method should be trained on the data used in the clustering, each data item labeled according to class. Finally the trained SVM-CD classifier can be used to classify new data.

## 6. Acknowledgments.

Diogo Ayres de Campos, MD, PhD is acknowledged for the coordination of data collection within the SisPorto multicentre study - PECS/C/SAU/207/95.

We also acknowledge projects Cardiocronos - POSI/CPS/ 40153/200 and KCrypt-POSC/EIA/60819/2004 and CINTESIS and LIACC research centers - Programa de Financiamento Plurianual, Fundação para a Ciência e Tecnologia and Programa POS, Portugal.